\documentclass[twocolumn,showpacs,aps,prl]{revtex4}
\usepackage{epsfig}

\begin{document}

\title{Spatial resolution of spin waves in an ultra-cold gas}
\author{J.~M. McGuirk$^\ast$, H.~J. Lewandowski, D.~M. Harber, T. Nikuni$^{\dag}$, J.~E. Williams$^{\ddag}$ and E.~A. Cornell$^\ast$}
\affiliation{JILA, National Institute of Standards and Technology and Department of Physics, \\
University of Colorado, Boulder, Colorado 80309-0440}
\date{\today}

\begin{abstract}
We present the first spatially resolved images of spin waves in a gas.  The complete longitudinal and transverse spin field as a function of time and space is reconstructed. Frequencies and damping rates for a standing-wave mode are extracted and compared with theory.
\end{abstract}

\maketitle

Macroscopic collective behavior can arise in a quantum gas even above the onset of degeneracy.  When indistinguishable atoms collide, the scattering can be significantly altered by the need to symmetrize the scattered waves.  Examples of quantum scattering effects are seen in the strong polarization dependence of heat conduction and viscosity coefficients in spin polarized $^3$He \cite{emery1964} and of thermalization rates in nondegenerate Fermi gases \cite{jin1999}.  Quantum collisional effects can also produce spatio-temporal spin oscillations known as spin waves.  In a recent experiment, we observed in an ultra-cold atomic gas a startling spin migration effect which we called ``anomalous spin-state segregation'' \cite{lewando2002}. Several theory groups have shown that this effect can be thought of as a half-cycle of a large-amplitude, over-damped spin wave \cite{newswtheory}.  In this Letter we verify experimentally their viewpoint and present spatially resolved images of coupled transverse and longitudinal spin perturbances propagating through a magnetically trapped atomic cloud. Frequencies and damping rates for the standing-wave excitations are studied as a function of density and temperature.

Although the concept of spin waves in ferromagnets dates back to Bloch's predictions in 1930 \cite{bloch1930}, the theory of spin waves in liquids and in dilute gases was not formulated until the much later \cite{spinwavetheory}.  The first evidence for spin waves (in the form of extra resonances in nuclear magnetic resonance spectra) occurred in spin polarized hydrogen and soon after in polarized $^3$He and dilute mixtures of $^3$He in $^4$He \cite{spinwaveexp}. On a microscopic level, spin waves arise from interference effects in lowest partial-wave collisions. When identical particles collide, the $180^\circ$ backscattering event causes a scattered atom to propagate along a trajectory indistinguishable from that of either a forward scattered atom or an unscattered atom. When the two atoms are spin aligned, the backward scattering contribution is indistinguishable from the forward scattering contribution, which adds a factor of two to the mean-field collisional energy shift. When the spins are antiparallel, the backward scattering event is distinguishable and the mean-field shift arises from the forward scattering event only. In the case of intermediate spin alignment, the backward scattering event is only partially distinguishable from the unscattered event, and one needs to add coherently two unparallel spin amplitudes, leading to a rotation of the spins of the scattered atoms. The cumulative effect of many such spin-rotating collisions is such that inhomogeneities of spin propagate like waves rather than diffusively. The characteristic frequency scale for the spin rotation effect is the exchange collision rate, $\omega_{\mbox{\scriptsize{exch}}} = 4 \pi \hbar a n/m$, where $m$, $n$, and $a$ are the mass of the atoms, the number density, and the s-wave scattering length respectively.

The experimental apparatus and spectroscopy method used to prepare and probe the ultra-cold gas are described in \cite{lewando2002} and briefly will be summarized here. $^{87}$Rb atoms are precooled and trapped in a magneto-optical trap, transferred to a Ioffe-Pritchard magnetic trap via a servo-controlled linear track, and cooled further using forced radio-frequency evaporation.  The trapping frequencies are \{7, 230, 230\}~Hz, and typical atom cloud parameters are $n \sim 10^{13}$~cm$^{-3}$ and $T \sim 600$~nK, several times the Bose-Einstein condensation temperature. The radial frequency is high enough that for the effects considered here it is sufficient to average the cloud over the radial dimensions, effectively reducing the cloud to a one dimensional density distribution.  There are two conveniently trapped atomic hyperfine sates, labelled $|1\rangle$ and $|2\rangle$, which together make up a pseudo-spin doublet \cite{states}. The spin state is manipulated using a two-photon coupling transition \cite{lewando2002}. Typical Rabi frequencies are around 3~kHz.

The transition frequency $\Delta(z)$ is in general a function of the axial position $z$, due to the spatial inhomogeneity of the Zeeman shift and of the density-dependent mean-field shift.  In Ref.~\cite{lewando2002} we describe in detail how the bias magnetic field can be tuned so as to cause the Zeeman shifts to cancel in whole or in part the mean-field shifts. As a consequence, the spatial curvature of the frequency, $\partial^2\!\Delta(z) /\partial z^2$, is a controllable parameter of the experiment which may be tuned in real time, even as a wave is propagating.

To describe the effective two-level system, we use the language of the Bloch sphere (see \cite{allen1987}) and take the axis \emph{w} as the ``longitudinal'' population inversion, and the axes \emph{u} and \emph{v} as the ``transverse'' coherences \cite{sxsz}. In a typical experiment, the atom-preparation cycle concludes with the atoms initially all in the $|1\rangle$ state, along the \emph{w}-axis. Applying a $\pi/2$ pulse starts the evolution of the wave by rotating all the spins to lie along the \emph{v}-axis. When $\Delta(z)$ is set to have nonzero curvature, the spins along the long axis of the cloud begin to fan out in the \emph{u-v} plane as the relative phase of the $(|1\rangle + |2\rangle)$ coherent superposition develops a spatial dependence. Optionally at this point in the evolution, we can tune $\partial^2\!\Delta(z) /\partial z^2$ back to zero or leave it unchanged and watch the spins evolve in an inhomogeneous potential. In either case, the atoms' thermal motion now carries spin information back and forth along the length of the cloud, and spin-rotating collisions can tilt the spins up out of the \emph{u-v} plane \cite{meanfield}.  Eventually, spin-currents develop that drive oscillations about an equilibrium spin distribution. Our driving potential, determined by the curvature of $\Delta(z)$, is even in $z$, and we predominantly drive the lowest order symmetric mode: a two-node standing wave.

\begin{figure}
\leavevmode
\epsfxsize=3.375in
\epsffile{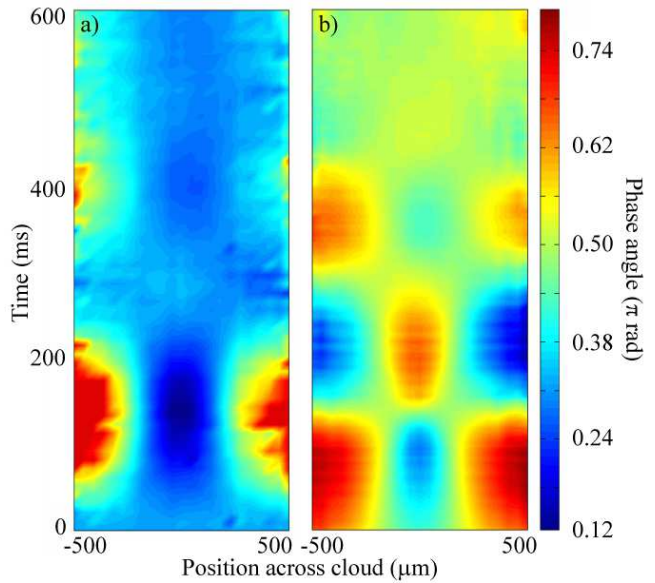} 
\caption{\label{fig:2d} False color plots of the a) longitudinal and b) transverse components of the Bloch vectors across the cloud as a function of time for a density of $2\times10^{13}$~cm$^{-3}$ and a temperature of 800~nK (corresponding to a $380\ \mu$m full width, half maximum cloud size). The angles have been interpolated between bins and time steps of a) 10~ms and b) 3~ms.}
\end{figure}

The longitudinal and transverse components of the Bloch vectors are measured as a function of time and space.  The method for projecting the longitudinal component is similar to that described in \cite{lewando2002}. After an initial resonant $\pi/2$ pulse, the Bloch vectors are allowed to evolve for times up to $t = 800$~ms, and then the atoms are imaged, thereby projecting the Bloch vectors along the cloud onto the two spin states $|1\rangle$ and $|2\rangle$.  The cloud is radially averaged and broken up into 25 equal bins along the axial direction for spatial resolution of the spin dynamics. The $|1\rangle$ and $|2\rangle$ populations are measured on separate experimental shots, and the longitudinal angle $\theta$ of the Bloch vector is extracted. ($\theta = \pi/2$ corresponds to equal populations, and $\theta = 0$ or $\pi$ correspond to all the atoms in $|1\rangle$ or $|2\rangle$ respectively.)  The evolution is traced out by repeating the experiment for many values of evolution time.  Fig.~\ref{fig:2d}a shows the time evolution of the longitudinal component of the spin across the cloud. The initial oscillation appears to be ``spin-state segregation'' \cite{lewando2002} but in fact is the first half-cycle of a spin wave as the Bloch vectors in the cloud center tip up from the transverse plane and the vectors near the cloud edge tip down.

The transverse spin measurement consists of a two-pulse $\pi/2 - \pi/2$ Ramsey pulse sequence separated by an evolution time $t$ \cite{ramsey1956}.  For each time evolution time $t$ and for each spatial bin along the trap axis, we measure the final density of the atoms in $|1\rangle$, resulting in a set of roughly sinusoidal (but with slightly time-varying phase and amplitude) Ramsey fringes for each location in space. The transverse phase and amplitude of the Bloch vector are extracted from the fringes at each time $t$ for each radially-averaged section of the cloud by performing a sinusoidal fit over a small window centered about $t$ in each spatial bin \cite{transverse}. Fig.~\ref{fig:2d}b shows typical phases extracted with this method. The spatial and temporal oscillations around the mean phase are the transverse projections of the spin waves.

\begin{figure*}
\leavevmode
\epsfxsize=7in
\epsffile{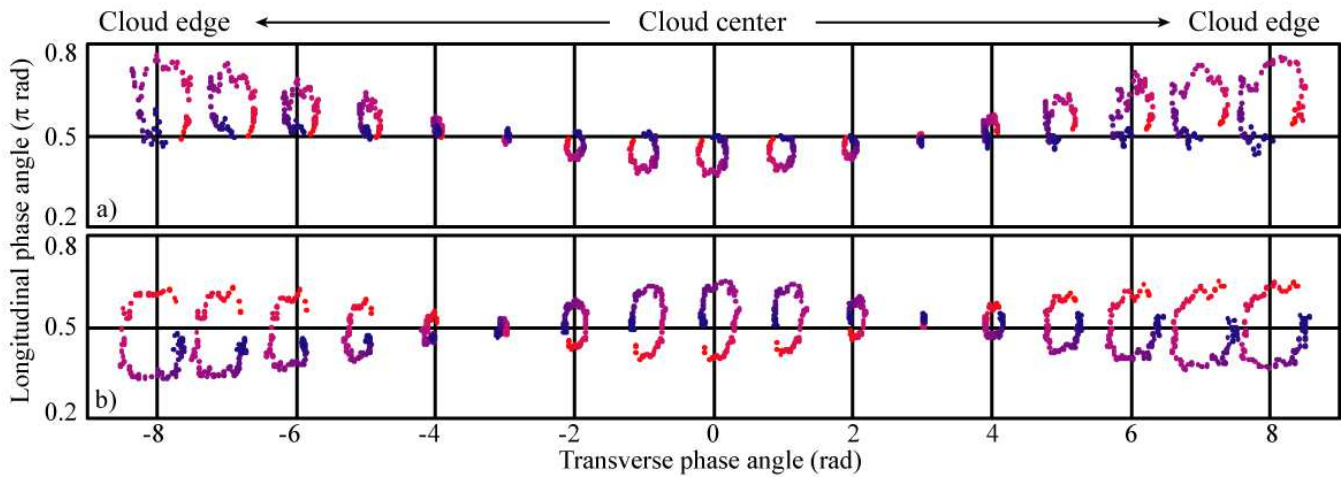} 
\caption{\label{fig:traj} Bloch vector trajectories across the cloud for $T = 600$~nK and $n = 2 \times 10^{13}$~cm$^{-3}$.  Each spatial bin ($\sim 40\ \mu$m/bin) is shifted by 1~rad from the adjacent bins (see text).  In each bin, the path traced by the Bloch vectors is shown starting with the red points near time $t = 0$ and spiralling towards the blue points at $t = 0.3$~s as the Bloch vectors precess about the equilibrium position.  a) Spin waves at $\partial^2\nu/\partial z^2 \approx$ 90~Hz/mm$^2$.  The equilibrium position (center of precession) has a curvature to match the curved potential. b) Spin waves are excited for 70~ms at the same curvature, and then at $t = 0$ the bias magnetic field is ramped quickly so that $\partial^2\nu/\partial z^2 \approx 0$, giving an approximately flat equilibrium configuration near the \emph{v}-axis for all spatial bins.}
\end{figure*}

The complete trajectories of the Bloch vector can be determined for each spatial bin by combining the transverse and longitudinal components.  Normalizing the length of each vector to the length at time $t = 0$ removes any effects of decoherence (\emph{e.g.} from inelastic loss and magnetic field fluctuations) and allows for a study of the equilibrium phases of the Bloch vectors as a function of the position across the cloud for various $\Delta(z)$.  Fig.~\ref{fig:traj} shows the trajectories of the reconstructed Bloch vectors in each spatial bin.  Each bin has been offset by 1~rad in the transverse direction from the adjacent bins for clarity and to depict schematically the position across the cloud.  The oscillations can be fit to extract spin wave frequency, amplitude, and damping rate.  

The steady-state Bloch vector arrangement is determined by the curvature of the potential.  In order to equilibrate, the Bloch vectors across the cloud must arrange themselves so that the torque from the driving potential $\Delta(z)$ (which tends to spread the vectors in the \emph{u-v} plane) is balanced by a torque from the local curvature in the spin field, which is spread out in the \emph{v-w} plane. In the presence of a nonuniform potential, the equilibrium configuration of the Bloch vectors is curved longitudinally across the cloud to match the curvature of $\Delta(z)$, \emph{i.e.} the Bloch vectors lie in the \emph{v-w} plane but are not all collinear with the \emph{v}-axis.  However, if the bias magnetic field is chosen to cancel the density shift, then $\Delta(z)$ is predominantly flat across the cloud, and the equilibrium corresponds to the Bloch vectors everywhere in the cloud collinear with the \emph{v}-axis.  These equilibrium configurations are illustrated in Fig.~\ref{fig:traj}a and Fig.~\ref{fig:traj}b, which show spin-wave evolution in curved and flat potentials respectively.

\begin{figure}
\leavevmode
\epsfxsize=3.375in
\epsffile{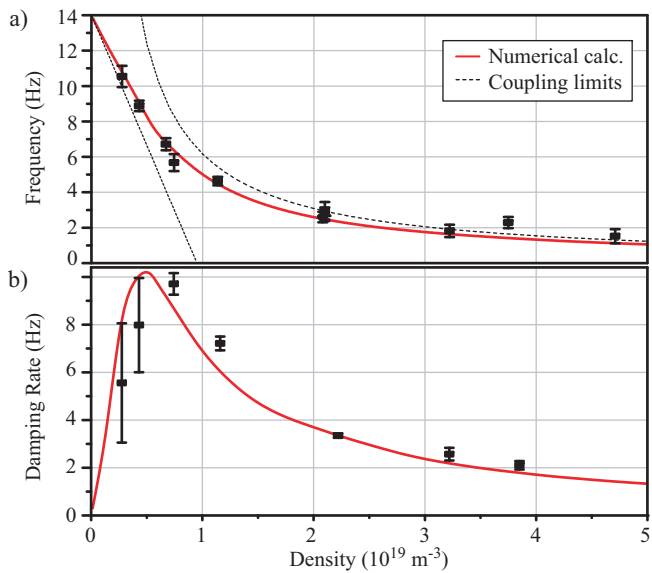} 
\caption{\label{fig:nodes} a) Frequency of the transverse spin wave component as a function of the peak density.  Each frequency is an extrapolation to the limit of small amplitude spin waves to remove the dependence on driving amplitude. The solid line is the predicted dispersion relation, and the dotted lines show the weak and strong coupling limits.  All data is for $T = 600$~nK except for the three tightly clustered points at $n = 2\times 10^{19}$~m$^{-3}$ which correspond to $T$ = 350, 600, and 1000~nK. b) Damping rate as a function of density, including a numerical calculation (solid line) which includes elastic collisions and Landau damping.}
\end{figure}

Next, the density dependence of the excitation frequency was studied using transverse spin waves.  The excitation frequency exhibits a strong density dependence, scaling roughly as $1/n$, shown in Fig.~\ref{fig:nodes}a. The linear regime for these excitations occurs only at extremely small amplitude, where spin waves are difficult to observe due to signal-to-noise limitations.  In the nonlinear regime there is a dependence of frequency on amplitude.  In order to remove this dependence, we excite spin waves over a range of amplitudes by controlling the curvature of the potential, $\partial^2 \Delta(z) / \partial z^2$, and we extrapolate the frequency to the low amplitude limit for each density.  The solid line is a numerical calculation predicting the frequency obtained from solving the one-dimensional Boltzmann transport equations \cite{williams2002b}.  The two dotted lines represent the weak and strong coupling limits, $\omega_{\mbox{\scriptsize{exch}}} \ll \omega_\circ$ and $\omega_{\mbox{\scriptsize{exch}}} \gg \omega_\circ$ respectively:
\begin{eqnarray}
\label{freq}
\omega_{\mbox{\scriptsize{w}}} &=& 2 \omega_\circ-\frac{\omega_{\mbox{\scriptsize{exch}}}}{2}\\
\label{freq2}
\omega_{\mbox{\scriptsize{s}}} &=& \frac{k_{\mbox{\scriptsize{eff}}}^2 (k_{\mbox{\scriptsize{b}}} T)}{m\,\omega_{\mbox{\scriptsize{exch}}}}.
\end{eqnarray}
where $\omega_\circ$ is the axial trap frequency and $k_{\mbox{\scriptsize{eff}}}$ is an effective wave vector.  The spatial wave vector $k = 2 \pi/\lambda$ of the spin waves is studied by finding the distance $\lambda/2$ between the spatial nodes, \emph{i.e.}~the bins where the orientation of the Bloch vector remains constant.  The wave vectors $k$ and $k_{\mbox{\scriptsize{eff}}}$ differ by a numerical factor of order unity due to the finite size of the sample.  For all the work discussed here, the lowest order symmetric mode, $\eta = 2$, is excited by the symmetric inhomogeneous potential \cite{mode}. The wavelength of the $\eta = 2$ mode is observed to be directly proportional to the size of the trap. As predicted by Eqs.~\ref{freq} and \ref{freq2}, there is no temperature dependence for the excitation frequency, since we observe $k$ to scale inversely as the size of the cloud, \emph{i.e.}~as $T^{-1/2}$. The numerical calculation agrees well with experiment (see Fig.~\ref{fig:nodes}a).

The Bloch vectors initially precess about the equilibrium value with some initial amplitude. As the Bloch vectors precess, their trajectories spiral towards the equilibrium configuration due to damping of the spin wave excitation.  The damping is predicted to be a combination of two effects: spin diffusion and Landau damping \cite{levitov1999}.  Spin diffusion dominates in the high density, or hydrodynamic, regime.  For intermediate densities, there is a maximum in the damping rate due to the combination of the two effects.  It near this intermediate regime that these experiments are performed. We observe a relationship between spin wave lifetime and density which is good agreement with theory (see Fig.~\ref{fig:nodes}b).

Our original motivation to explore spin dynamics in uncondensed clouds arose from our plan to extend to finite temperatures our earlier studies of spin coherence in a condensate \cite{hall1998b} and in particular to understand the extent to which an unpolarized normal cloud acts as a decohering thermal reservoir. The existence of spin waves will complicate these studies. A pure condensate cannot support spin waves, but it will trivially separate due to small differences in the scattering lengths \cite{hall1998}. Spin oscillations in the thermal cloud can have a considerable effect on the spin dynamics of a finite temperature condensate, and conversely, having a small stationary cluster of condensed atoms will significantly perturb spin waves in a thermal cloud.  Such a two component system is a rich system for studying spin dynamics in ultra-cold gases \cite{levitov1999}.

In conclusion, we have observed collective spin excitations in a nondegenerate ultra-cold gas.  Particle indistinguishability and the need to symmetrize collisions between like particles in coherent superpositions give rise to spatial and temporal oscillations of the Bloch vectors across a cloud of trapped atoms. We have separately imaged both transverse and longitudinal components of these spin waves and studied the effects of temperature and density on the spin wave frequency, paving the way for studying spin dynamics in a finite-temperature condensate system.

We acknowledge useful conversations with the other members of the JILA BEC collaboration. This work was supported by grants from the NSF and NIST.


\begin{thebibliography}{21}
\bibitem{states}
The two states in question are $|1\rangle \equiv |F = 1, m_{\mbox{\scriptsize{F}}} = -1 \rangle$ and $|2\rangle \equiv |F = 2, m_{\mbox{\scriptsize{F}}} = 1 \rangle$, which are pseudo-spin states since the system is not a true spin-$1/2$ system.
\bibitem{meanfield}
It is important to realize that the difference in mean-field interactions between the two states, $(a_{22}-a_{11}) \approx 2.6$~\AA, is extremely small in $^{87}$Rb and is not the mechanism for spin wave propagation.  For the purposes of these experiments $(a_{22}-a_{11})$ provides merely a contribution to the inhomogeneity in $\Delta(z)$.  The total mean-field effect $\omega_{\mbox{\scriptsize{exch}}}$ is proportional to $a = (a_{11}+a_{22})/2 \approx 50$~\AA.
\bibitem{transverse}
The mean phase of the Bloch vector evolves at the average hyperfine splitting frequency of $\overline{\Delta} = 6.8$~GHz; the Bloch vector only moves slowly in the frame rotating at $\overline{\Delta}$.  It is in this reference frame the the effects of spin waves are considered. The local oscillator is set so that the Ramsey fringe frequency is much larger than the spin wave frequency, simplifying analysis of the ``instantaneous'' phase and amplitude of the Bloch vectors.
\bibitem{mode}
The quadrupole mode is excited with a frequency splitting $\Delta(z)$ that varies quadratically with distance from the trap center.  The dipole mode could be driven by applying a linear gradient to the trapped atoms.
\end{thebibliography}
\end{document}